\documentclass[10pt,a4paper,showpacs,twocolumn]{revtex4}
\usepackage{amssymb}
\usepackage{natbib}
\usepackage{graphicx}
\begin{document}
\title{ Higher-Order
Nonlinear Schr\"{o}dinger equation with derivative non-Kerr nonlinear terms: A model for sub-10fs pulse propagation}
\author{Amitava Choudhuri$^a$ and K Porsezian$^b$}
\email{amitava_ch26@yahoo.com(A. Choudhuri)}
\affiliation{$^a$Department of Astrophysics, School of Physical, Chemical and Applied Sciences, Pondicherry University, Pondicherry 605014, India}
\affiliation{$^b$Department of Physics, School of Physical, Chemical and Applied Sciences, Pondicherry University, Pondicherry 605014, India}
\pacs{42.81.Dp, 05.45.Yv, 42.65.Tg, 42.79.Sz}
\begin{abstract}
We analytically solved the higher-order nonlinear
Schr\"{o}dinger (HNLS) equation with non-Kerr nonlinearity under
some parametric conditions and investigated explicitly bright and
dark solitary wave solutions. Periodic wave solutions are also
presented. The functional form of the bright and dark
solitons presented are different from fundamental known $sech(.)$ and
$tanh(.)$ respectively. We have estimated theoretically the size of the derivative non-Kerr nonlinear coefficients of the HNLS equation that agreed the reality of the waveguide made of highly nonlinear optical materials, could be used as the model parameters for sub-10fs pulse propagation.
\end{abstract}
\maketitle Optical solitons  have promising potential to become
principal information carriers in telecommunication due to their
capability of propagating long distance without attenuation and
changing their shapes. Considerable attentions are being paid
theoretically and experimentally to analyze  the dynamics of
optical solitons in optical waveguide. The waveguides used in the
picosecond optical pulse propagation in nonlinear optical
communication systems are usually of Kerr type and consequently
the dynamics of light pulses are described by nonlinear
Schr{\"o}dinger (NLS) family of equations with cubic nonlinear
terms \cite{HT}. The validity of the NLS equation as a reliable model is
dependent on the assumption that the spacial width of the
soliton is much larger than the carrier wavelength. This
is equivalent to the condition that the width of the soliton
frequency spectrum is much less than the carrier frequency. The robustness of the optical soliton makes it
useful for long distance optical communication systems,
the high frequency of the optical carrier makes possible
a high bit rate, and to increase the bit rate further it is
desirable to use shorter femtosecond pulses and in order to model the propagation of a femtosecond ($<$100fs)
optical pulse in an optical fiber, higher order nonlinear Schr\"{o}dinger
equation (HNLS) (not including optical fiber
loss)   \cite{HK}
$$E_z=i(a_1E_{tt}+a_2|E|^2E)+a_3E_{ttt}+a_4(|E|^2E)_t
+a_5E(|E|^2)_t\,\,\eqno(1)$$ is required. Here $z$ is the normalized distance
along the fiber, $t$ is the normalized time with the frame of the
reference moving along the fiber at the group velocity. The
subscripts $z$ and $t$ denotes the spatial and temporal partial
derivatives respectively. The coefficients
$a_i\,(i=1,\,2,\,...,\,5)$, particularly, $(a_1={\beta_2\over 2}, \,\,
a_2=\gamma_1,\,\,a_3={\beta_3\over 6},\,\,a_4=-\frac{\gamma_1}{\omega_0}
\,\,{\rm and}\,\,a_5=\gamma_1\,T_R)$ are the real parameters related to
group velocity dispersion (GVD), self phase modulation (SPM),
third-order dispersion (TOD), self steepening and self-frequency
shift due to stimulated Raman scattering (SRS) respectively. Here
$\beta_j=(\frac{d^j\beta}{d\omega^j})_{\omega=\omega_0}$ is the
dispersion coefficients evaluated at the carrier frequency $\omega_0$,
with $\beta_1$, the inverse of group velocity, $\beta_2$, the group
velocity dispersion parameter, $\beta_3$  third order dispersion
(TOD) parameter and so on. $\beta$ is propagation constant. More specifically, $\gamma_1$ is coefficient of cubic nonlinearity, which results from the intensity dependent refractive index. The term related to $\frac{\gamma_1}{\omega_0}$ results from the intensity dependence of the group velocity and causes self steepening and shock formation at the pulse edge. The last term related to $a_5=\gamma_1\,T_R$ incorporates the intrapulse Raman scattering and originates from the delayed response, which cause a self-frequency shift, where $T_R$, is called Raman time constant, can be estimated from the slope of the Raman gain (SRS). The characteristic Raman time constant $T_R$ is defined as the first moment of the nonlinear response function \cite{GPA}. Actually, to model intrapulse Raman Scattering the last coefficient $a_5$ should be $a_5=i\gamma_1\,T_R$. However, in the present work we succeeded in deriving analytical solutions in the case when $a_5$ was real. Let us remark that the latter case ($a_5$ real) also dominates in the analytical studies (e.g., Painlev{\'e} property, inverse scattering transform, Hirota direct method, conservation laws ) undertaken to date \cite{PN} to show its integrable nature and obtained different types of exact
solutions such as new solitary wave solution, w-shaped solution,
bright and dark optical solitary wave solutions etc. Thus besides considering that the present works \cite{PN} realizes a significant advance as compared to previous literature, extension of the work to the case of $a_5$ imaginary represents a theoretical challenge that should be undertaken in the near future.\par Present days
applications in telecommunication and ultrafast signal routing
systems, as the intensity of the incident light field becomes
stronger, non-Kerr nonlinear effects come into play and due to
these additional effects, the physical features  and the stability
of the NLS soliton can change.  The way through the non-Kerr
nonlinearity influences NLS soliton propagation is described by
the NLS family of equations with higher degree of nonlinear terms
\cite{RKL}.  To increase the channel handling capacity and ultra high speed pulse, it is necessary to transmit solitary waves at a high bit rate ($\approx 1-10 \,fs$) of ultrashort pulses, which can be seen in many
applicative contexts such as high repetition pulse sources based on fiber technology \cite{AUTO}. At the same time, it is also important to include some additional higher-order perturbation effects to HNLS equation to analyze the solitary wave solution in non-Kerr nonlinear medium.
\par Here, in this Letter,  we consider the higher-order NLS (HNLS)
equation with non-Kerr term \cite{ACKP}, can be written in terms
of slowly varying complex envelope of the electric field $E(z,t)$,
as {\small$$E_z=i(a_1E_{tt}+a_2|E|^2E)+a_3E_{ttt}+a_4(|E|^2E)_t
+a_5E(|E|^2)_t+$$$$ia_6|E|^4E+a_7(|E|^4E)_t+
a_8E(|E|^4)_t\,\,.\eqno(2)$$} The terms related to coefficients
$a_6,\,a_7,\,a_8$ in Eq. $(2)$ represent the quintic non-Kerr
nonlinearities.  The quintic nonlinearities arise from the expansion
of the refractive index in power of intensity $I$ of the light
pulse : $n=n_0+n_2\,I+n_4\,I^2+...$. Here $n_0$ is the linear
refractive index coefficient and $n_2$, $n_4$ are the nonlinear
refractive index coefficients, originate from third- and
fifth-order susceptibilities respectively. The polarizations induced through
these susceptibilities give the cubic and quintic (non-Kerr) terms
in nonlinear Schr\"{o}dinger equation respectively. The
nonlinearity arises due to fifth-order susceptibility can be
obtained in many optical materials such as semiconductors,
semiconductor doped glasses, AlGaAs, polydiacctylene toluene sulfonate
(PTS), chalcogenide glasses and some transparent organic materials.
When the last three terms related to $a_6,\,a_7,\,a_8$  of Eq.
$(2)$ are ignored, the resulting equation becomes the HNLS
equation as given in Eq. $(1)$.   In a recent paper \cite{ACKP} we
have investigated the Dark-in-the-Bright (DITB) solitary wave
solution of Eq. $(2)$. The DS or DITB solitary wave solution is composed of the product of bright and dark solitary waves.  We
also investigated the stability of the DITB solution under some
initial perturbation on the parametric conditions. The shape of
pulse remains unchanged up to $20$ normalized length even under
some very small violation in parametric conditions. More recently
\cite{AK}, we have studied modulational instability (MI) of Eq.
$(2)$ with forth-order dispersion in context of optics and presented an analytical expression for
MI gain to show the effects of non-Kerr nonlinearities
and higher-order dispersions on MI gain spectra. In our
study we also demonstrate that MI can exist not only for
anomalous group velocity dispersion (GVD) regime but
also in the normal GVD regime and also investigated that the quintic non-Kerr nonlinear terms are more important over the cubic Kerr
nonlinearity because non-Kerr nonlinearities are responsible for
stability of localized solutions. But in the previous works \cite{RKL,ACKP,AK} the solitary wave solutions in presence of higher non-Kerr nonlinearity have not been investigated. In a very recent work, Triki and Taha \cite{TT} presented solitary wave solutions for HNLS equation including non-Kerr nonlinear terms upto the coefficient $a_6$ of Eq. $(2)$.
In this article, we shall study the bright- and dark solitary wave solutions for HNLS equation that contains time derivative of non-Kerr nonlinear terms and estimate the size of model coefficients of Eq. $(2)$  which will be useful for propagation of very short pulse of width around sub-10fs in highly nonlinear optical fibers.
\par To investigate the existence of
analytic wave solution of HNLS equation in presence of non-Kerr
terms we begin by scaling the variables of the Eq. $(2)$ in the
form
$$E=b_1\Psi\,\,,\,\,\,\,\,z=b_2\xi\,\,,\,\,\,\,\,\, {\rm and} \,\,\,t=b_3\tau$$
and choosing $b_1$, $b_2$ and $b_3$ such that the coefficients
corresponding to GVD, SPM and TOD become unity. Thus in the scaled
formed of the Eq. $(2)$ becomes
{\small$$\Psi_\xi=i(\Psi_{\tau\tau}+|\Psi|^2\Psi)+\Psi_{\tau\tau\tau}+\alpha_1(|\Psi|^2\Psi)_\tau+
\alpha_2\Psi(|\Psi|^2)_\tau+$$$$i
\alpha_3|\Psi|^4\Psi+\alpha_4(|\Psi|^4\Psi)_\tau+\alpha_5\Psi(|\Psi|^4)_\tau\eqno(3)$$}
where
$$\alpha_1=\frac{b_1^2b_2a_4}{b_3}=\frac{a_4a_1}{a_2a_3}\,\,,\,\,\,\alpha_2=\frac{b_1^2b_2a_5}{b_3}=
\frac{a_5a_1}{a_2a_3}\,\,,$$$$
\alpha_3=b_1^4b_2a_6=\frac{a_6a_1^3}{a_2^2a_3^2}\,\,,\,\,\,\alpha_4=\frac{b_1^4b_2a_7}{b_3}=
\frac{a_7a_1^4}{a_2^2a_3^3}\,\,\,{\rm and}
\,\,\,$$$$\alpha_5=\frac{b_1^4b_2a_8}{b_3}=\frac{a_8a_1^4}{a_2^2a_3^3}\,\,.$$
In writing $(3)$ we have chosen
$b_1=\left(\frac{a_1^3}{a_2^2a_3^2}\right)^{1\over2}$,
$b_2=\frac{a_3^2}{a_1^3}$ and $b_3=\frac{a_3}{a_1}$.\par To obtain
the exact solitary wave solutions of Eq.(3) we consider a solution
 of the  following form
$$\Psi(\xi,\,\tau)={\cal P}(\tau+v \xi)e^{i(k \xi-\Omega
\tau)}={\cal P}(\chi)e^{i(k \xi-\Omega \tau)}\,\,,\eqno(4)$$ with
${\cal P}(\chi)$ real. On substitution Eq. (4) into Eq. (3) and
 after removing the exponential terms one can obtain the real and imaginary parts
of the resulting equation as {\small$${\cal
P}_{\chi\chi}=(v-2\Omega+3\Omega^2){\cal P}-\frac{3\alpha_1+
2\alpha_2}{3}{\cal P}^3-\frac{5\alpha_4+4\alpha_5}{5}{\cal
P}^5\eqno(4a)$$} and {\small$${\cal
P}_{\chi\chi}=\frac{(k+\Omega^2-\Omega^3)}{(1-3\Omega)}{\cal
P}-\frac{(1-\Omega\alpha_1)}{(1-3\Omega)}{\cal P}^3
-\frac{(\alpha_3-\Omega\alpha_4)}{(1-3\Omega)}{\cal
P}^5\,\,.\eqno(4b)$$} In the following we will discuss two cases:\\
\\{\bf Case $1$: $\Omega\neq\frac{1}{3}$.} Equating these two
equations we get the following necessary and sufficient conditions
on $\Omega$ and equation for $k$ in terms of $\Omega$:
$$\Omega=\frac{3\alpha_1+2\alpha_2-3}{6(\alpha_1+\alpha_2)}=
\frac{5\alpha_4+4\alpha_5-5\alpha_3}{(10\alpha_4+12\alpha_5)}\,\,\eqno(5a)$$
and
$$k=(1-3\Omega)(v-2 \Omega+3\Omega^2)-\Omega^2+\Omega^3\,\,.\eqno(5b)$$
with constraint relations
$$\alpha_4=\frac{3\alpha_1}{5}\,\,\,,\,\,\,\,\,\,\,
\alpha_5=\frac{\alpha_2}{2}\,\,\,\,\,\,\,\, {\rm and}
\,\,\,\,\,\,\,\,\,\alpha_3=\frac{3}{5}\,\,.\eqno(6)$$ The function
${\cal P}(\chi)$ satisfies the ordinary nonlinear differential
equation in $(4a)$. Multiplying $(4a)$ by ${\cal P}_\chi$ and
integrating we get
$$({\cal P}_{\chi})^2=a{\cal
P}^2-b{\cal P}^4-c{\cal P}^6+2 {\cal E}\,\,,\eqno(7)$$ where
$(v-2 \Omega+ 3\Omega^2)=a$,
$\frac{3\alpha_1+2\alpha_2}{6}=b$,
 $\frac{5\alpha_4+4\alpha_5}{15}=c$ and ${\cal E}$ is the arbitrary constant of integration. Eq. $(7)$ describe the evolution of the anharmonic oscillator with potential
$$U({\cal P})=-\frac{a}{2}{\cal P}^2+\frac{b}{2}{\cal P}^4+\frac{c}{2}{\cal
P}^6\,\,\eqno(8)$$ and ${\cal E}$, the integration constant in
$(7)$ is the  energy of that anharmonic oscillator. For zero
energy $({\cal E}=0)$ we have find the solution of Eq.$(7)$ as
{\small$${\cal P}(\chi)=\frac{2 a^{\frac{3}{ 4}}\sqrt{(\sqrt{a}-b
\,e^{2\sqrt{a}\chi})^2+4ac\,e^{4\sqrt{a}\chi}}\,e^{\sqrt{a}\chi}}
{\sqrt{((a-(b^2-4ac)\,e^{4\sqrt{a}\chi})^2+16ab^2c\,e^{8\sqrt{a}\chi}}}\,\,,\eqno(9)$$}
provided that $a>0\,\,,\,\,\,b>0$. Now using Eqs.(4), (5b) and (9) we can write the solution of Eq.
$(3)$ as
$$\Psi(\xi,\,\tau)={\cal P}(\tau+v \xi)\,e^{i\left(((1-3\Omega)(v-2
\Omega+3\Omega^2)-\Omega^2+\Omega^3)
\xi-\Omega\tau\right)}\,\,.\eqno(10)$$
The intensity profile of the solitary wave solution (Eq. $(10)$)
is shown in Fig. $1(a)$, as computed from Eq. $(3)$ for the values
$\alpha_1=\alpha_2=1$ and $v=1$. One can check the
evolution of the intensity profile and it is an interesting to note that the wave profile remains unchanged during the evolution.
\begin{figure}[]
\centering
\raisebox{0.8in}{($a$)}\includegraphics[scale=0.5]{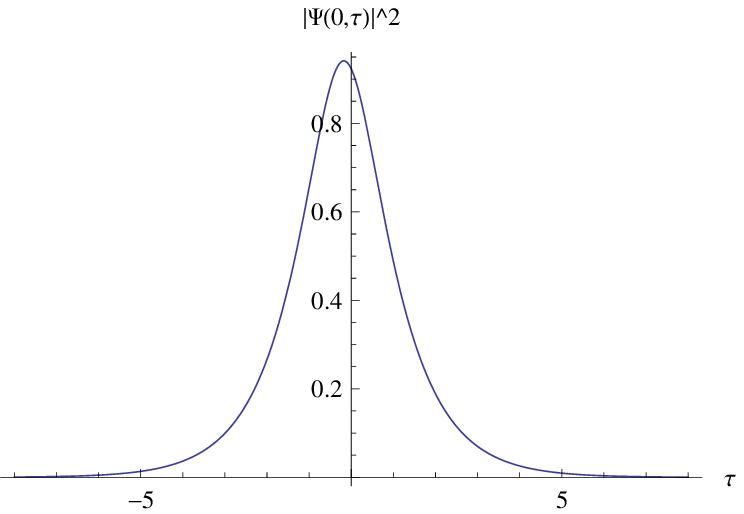}
\hskip 0.03cm
\raisebox{0.8in}{($b$)}{\includegraphics[scale=0.5]{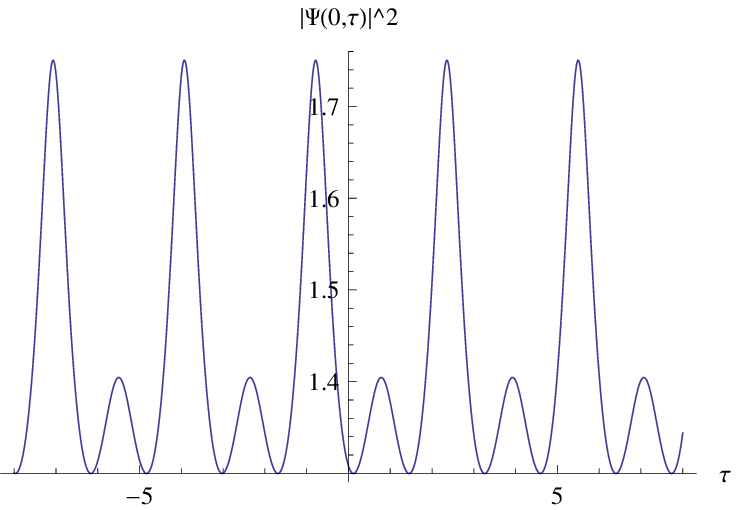}}
\caption{{\small (Color online) (a)Intensity of the solitary wave profile $|\Psi(0,\,\tau)|^2$
    as a function of $\tau$ for the value $\alpha_1=\alpha_2=1$ and $v=1$ (b) Periodic wave profile $|\Psi(0,\,\tau)|$
    as a function of $\tau$ as
    computed from Eq. $(10)$ for the value $\alpha_1=\alpha_2=1$ and $v=-1$. }}\label{Hpositon}
\end{figure}
For negative value of the parameter $a$, Eq. $(3)$ shows the
periodic solution. We have presented in Fig. $1(b)$  the
periodic wave solution. Here we
have taken the same parameter values as that in Fig. $1(a)$ but
$v=-1$ such that $a<0$. For $\Omega=0$ (i.e.
$3\alpha_1+2\alpha_2=3$) and consequently, $a=v$,
$b=\frac{1}{2}$ and $c=\frac{1}{5}$ one can verify the solitary
wave solution and periodic solution of Eq. $(3)$ from Eq. $(10)$
for $a>0$ and $a<0$ respectively.
\\ {\bf Dark solitary wave.} We have seen for ${\cal E}=0$, HNLS
equation in presence of non-Kerr terms supports the bight optical
wave solution provided that the parameter $a$ in Eq. $(7)$ is
positive. Now we chose ${\cal E}=-\frac{a^2}{6b}$ and
$c=-\frac{b^2}{3a}$ such that we can write Eq.$(7)$ in the form
$$d\chi=\left(-\frac{(a-b {\cal P}^2)^3}{3ab}\right)^{-\frac{1}{2}}d{\cal
P}\,\,.\eqno(11)$$ Integrating Eq. $(11)$ we find the solitary
wave solution of Eq. $(3)$, with the same constraint relations
stated above in $(6)$, as
$$\Psi(\xi,\,\tau)=\frac{a(\tau+v\xi)}{\sqrt{b}\sqrt{(-3+a(\tau+v
\xi)^2)}}\,e^{i\left(k\xi-\Omega\tau\right)}\,\,,\eqno(12)$$
with $k=(1-3\Omega)(v-2\Omega+3\Omega^2)-\Omega^2+\Omega^3$,
provided $a<0$, with $v=2
\Omega-3\Omega^2-\frac{5}{36}(3\alpha_1+2\alpha_2)$.
\begin{figure}
\centering
\includegraphics[scale=0.5]{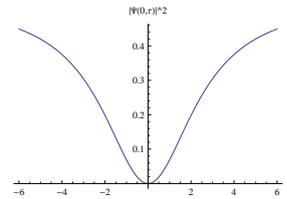}
\caption{\small (Color online) Intensity of the solitary wave profile $|\Psi(0,\,\tau)|$
(Eq.$(12)$) as a function of $\tau$ as computed from Eq. $(3)$ for
the value $\alpha_1=\alpha_2=1$ and $v=-\frac{4}{9}$}
\end{figure}
In Fig. $3$ we have plotted the intensity profile of the optical
 solitary wave solution with the parameter values
$\alpha_1=\alpha_2=1$ and $v=-\frac{4}{9}$. We call this
solution as a dark solitary wave in the sense that the intensity
profile associated with such soliton exhibits a dip in a uniform
background and asymptotic absolute value of $\Psi(\xi,\,\tau)$
tends toward a constant nonzero value for large values of
$\tau$. In this context, although bright solitons are relatively easy to generate in optical fiber, the dark solitons are less sensitive to optical fiber loss, less influenced by noise and are more stable against Gordon Haus jitter in long communication line \cite{KIV}. As the mutual interaction between two neighboring dark solitons is much weaker than that between two bright solitons \cite{Zhao}, so the properties of the dark soliton attracted scientists very much in the communication systems. But it is difficult to use a dark soliton with a $tanh(.)$-type wave form in a transmission system because dark pulse cannot be easily generated.  \\
\\{\bf Case $2$: $\Omega=\frac{1}{3}\,,$}
     Eq. $(4b)$ takes the form $$(k+\Omega^2-\Omega^3){\cal
P}-(1-\Omega\alpha_1){\cal P}^3 -(\alpha_3-\Omega\alpha_4){\cal
P}^5=0\,\,.\eqno(13)$$ Setting the coefficients of ${\cal P}$,
${\cal P}^3$ and ${\cal P}^5$ to zero in Eq. $(13)$ and using
$\Omega=\frac{1}{3}$ we obtain $k=-\frac{2}{27}$ and the constraint
condition $$\Omega=\frac{1}{\alpha_1}\,\,\,\,\,{\rm and}
\,\,\,\,\,\,\,\Omega=\frac{\alpha_3}{\alpha_4}, {\rm which\,\,
implies}\,\,\,\alpha_4=\alpha_1\,\alpha_3\,\,.\eqno(14)$$ In this
case, ${\cal P}{(\chi)}$ satisfy the ordinary differential
equation similar to that in $(7)$ but only differs in the
coefficient values.$$({\cal P}_{\chi})^2=a^\prime{\cal
P}^2-b^\prime{\cal P}^4-c^\prime{\cal P}^6+2 {\cal
E}^\prime\,\,.\eqno(15)$$ Here, in this case,
$(v-\frac{1}{3})=a^\prime$, $\frac{9+2\alpha_2}{6}=b^\prime$,
 $\frac{5\alpha_4+4\alpha_5}{15}=c^\prime$ and ${\cal E}^\prime$ is the arbitrary constant
of integration. In case $2$, similar to Eq. $(7)$ one can solve
Eq. $(15)$ to get the bright, dark and also periodic wave
solutions of Eq. $(3)$. Only here the coefficients $a^\prime$,
$b^\prime$ and $c^\prime$ are different from that in Eq. $(7)$.
\par Before conclusion, let us discuss some application of the above theoretical prediction. For large channel handling capacity in the frame of dense time-domain multiplexing and for high speed, it is necessary to transmit solitary wave at a high bit rate of ultrashort pulses. At the same time, it is also important to consider the higher-order non-Kerr like nonlinearity including derivative in HNLS equation for sub-10\,fs pulse propagation. The relevance of these terms is also important in the frame of post-soliton compression that can be achieved in highly nonlinear material.  Compared to silica glasses, as chalcogenide glasses exhibit an extremely high nonlinear refractive-index coefficient that can be two or three orders of magnitude greater than that of silica at 1.55 $\mu$m. They also offer several distinctive optical properties such as a transmission window that extends far into the infrared (IR) spectral region (up to 25 $\mu$m for telluride glasses). Because of high nonlinearity and large IR transparency, chalcogenide fibers are well suited for compact Raman amplifiers, supercontinuum generation and other mid-IR sources. For experimental verification of the propagation of solitary wave, one may use the  waveguide made of chalcogenide glasses which are made from heavy chalcogen elements such as S, As, Se, Te, having electron shells which are easily polarizable under an electro-magnetic field excitation.  For example,  the chalcogenide glass $As_2Se_3$,  the nonlinear index coefficients \cite{CHAL,UNG} are $n_2=2.2\times10^{-17}\,m^2/W$ and $n_4=-6.5\times10^{-31}\,m^4/W^2$ and for the $As_2S_3$ sample, the values are  $n_2=4.2\times10^{-18}\,m^2/W$ and $n_4=-6.0\times10^{-32}\,m^4/W^2$.
In general, the nonlinear coefficients $\gamma_i$ ($i=1,\,2$) can be estimated from $\gamma_i=\frac{2\pi n_j}{\lambda A^i_{eff}}$ , where $n_j\,(j=2,\,4)$ are the nonlinear refractive index coefficient. $A_{eff}=\pi w^2$ is the effective fiber core area, with $w$ is the core radius of the fiber, which varies from $(3-3.8)$ $\mu\,m$ for $As_2Se_3$ and $(1.3-1.7)$ $\mu\,m$ for the $As_2S_3$. $\lambda$, the typical telecommunication wavelength $1.55\mu m$.
For the range of $|\gamma_1|=(2000-3000)\,W^{-1}/Km$, cubic nonlinear coefficient, we have calculated the values for quintic nonlinear coefficients: $|\gamma_2|=(1.3-3)\,W^{-2}/Km$ for $As_2Se_3$ and for $|\gamma_2|=(3.3-7.6)\,W^{-2}/Km$$ As_2S_3$ respectively.
The $1.55\mu m$ window today is mainly of interest to long-distance telecommunications application. The other important feature of this transmitted wavelength is that it matches the fiber's low-loss regions. Fiber energy loss (absorption) compensation with sufficient Raman gain and distortionless propagation of picosecond soliton pulses in a monomodal optical fiber have been experimentally demonstrated by Mollenauer et. al. \cite{Mollen}. Chalcogenide glasses have attracted much interest in the past few years as a nonlinear optical material in the telecommunications wavelength window of 1550 nm, and are promising candidates for planar non-linear optical (NLO) rib waveguide devices due to  high nonlinearity, high refractive index, and non-linear optical losses (0.05 dB/cm) at 1550 nm \cite{JAP, Madden}. In a vary recent work, M. El-Amraoui et. al. \cite{OEXP} measured the fiber losses as low as 0.35 dB/m at 1.55µm for a 45 meters long 2.3 µm core size fiber. The related nonlinear Kerr coefficient is estimated as high as 2750 $W^{-1} km^{-1}$. It is also an important to note the fact that chalcogenide glasses exhibit the highest nonlinear refractive indices and suffer, at worst, only moderately from two-photon absorption, also they do not suffer from free-carrier absorption. The nonlinear absorption faced by the fiber material can be compensated using  derivative higher-order nonlinear Raman gain terms. In this context, Tuniz et. al. \cite{Tun} studied how Raman gain and nonlinear absorption counteract across the C and L-bands in two-photon absorption effects in single-mode chalcogenide fiber. \par Now, physically, for the ultrashort laser pulse propagation through optical fiber at telecommunication wavelength $1.55\,\mu m$, (last reference of \cite{HK}) and carrier frequency $\omega_0=1.22\times10^{15}s^{-1}$ i.e. $T_0=5.1475\times10^{-15}s$, if we choose the typical real experimental value for the model parameters of Eq. $(2)$ as $a_1=\frac{\beta_2}{2}=10\, ps^2/km$, $a_2=\gamma_1=2765\,W^{-1}/km$, $a_3=\frac{\beta_3}{6}=0.0235\, ps^3/km$, $a_4=-\frac{\gamma_1}{\omega_0}=-14.2328\,W^{-1}/((2\pi)km\,THz)$ and $a_5=\gamma_1T_R=14931\, W^{-1}\,fs/km\,\,(T_R=5.4\,fs$ for chalcogenide glass fiber \cite{UNG}), we can estimate the size of the coefficients of the non-Kerr nonlinearities of Eq. $(2)$ from the constraint relations in Eq.$(6)$. The calculated values for the coefficients of non-Kerr nonlinearities are $a_6=\gamma_2=\frac{3 a_2^2a_3^2}{5a_1^3}=2.533\,W^{-2}/km$, $a_7= \frac{3 a_2a_3^2a_4}{5a_1^3}=-1.304\times10^{-2}\,W^{-2}/((2\pi)km\,THz)$ and $a_8= \frac{ a_2a_3^2a_5}{2a_1^3}=11.3996\,W^{-2}\,fs/km$. For the sub-10fs bright pulse communication in non-Kerr medium  with the above model parameter values for Eq. $(2)$, one can check that we need chalcogenide optical fiber of the type $As_2Se_3$ of core radius $3.20415\mu m$ and for $As_2S_3$, the core radius is $1.399 \mu m$ respectively. If we use the same parameter values for the dark pulse propagation, one can obtain the energy value for dark pulse will be ${\cal E}=-\frac{a^2}{6 b}=-\frac{(v-2 \Omega+ 3\Omega^2)^2}{3\alpha_1+2\alpha_2}=-88.779$.
\par In conclusion, we have reported optical bright and dark solitary wave solutions of Higher-Order Nonlinear Schr\"{o}dinger
equation in presence of non-Kerr terms subject to constraint
relations among the parameters. The derivative Kerr and non Kerr nonlinear terms are important for compensation of the nonlinear absorption during propagation in highly nonlinear material. These terms also play an important role for the post-soliton compression to get highly stable compressed optical pulse. The functional forms of bright and dark solitary wave profiles reported are new. We also have
presented the periodic solutions which are very meaningful in
optics. We have seen that the estimated values for non-Kerr nonlinear model parameters of Eq. $(2)$ agreed the reality of the waveguide made of chalcogenide glasses. So the inclusion of the non-Kerr nonlinear terms in Eq. $(1)$ is justified to describe the sub-10fs pulse propagation in highly nonlinear optical fiber. As well as Case 1, we  have checked the parametric condition in Eq. $(14)$ for Case 2 using the model parameters given above. The quintic non-Kerr nonlinear terms in contemporary optics become very crucial to the upcoming applications in ultrafast signal routing systems, double doped optical fiber, optical switching etc. The periodic solution can be used to study the formation of solitons in the periodic stricture if one consider the quintic non-Kerr nonlinearity in fiber Bragg grating \cite{FBG}, and finally, with the calculated parameter values for highly nonlinear optical fiber made of chalcogenide glasses, the HNLS equation, in presence of non-Kerr nonlinear terms as higher-order perturbation, not only could find application in broadband telecommunication that extends far into the infrared (IR) spectral region for the bright and dark optical pulses but also Eq. $(2)$ may be a new theoretical model equation for experimental designing sub-10fs optical pulse propagation, which will be applicable for the next generation optical fiber using chalcogenide type high nonlinear optical glasses.
\\
\\
{\bf Acknowledgement}. The author (AC) is deeply indebted to Dr. Philippe Grelu, Professor of Physics and Optoelectronics at University de Bourgogne (Dijon, France) for constructive discussions.

\end{document}